\documentclass[reprint,twocolumn,amsmath,amssymb,aps,superscriptaddress,prb]{revtex4-1}
\usepackage{graphicx}
\usepackage{bm}
\usepackage{epsfig}
\usepackage{amssymb}
\usepackage{amsfonts}
\usepackage{amsmath}
\usepackage{color}
\usepackage{xcolor}
\usepackage{epstopdf}
\usepackage{hyperref}
\usepackage{float}
\usepackage{appendix}
\restylefloat{table}
\usepackage{bibentry}
\usepackage{multirow}
\usepackage[caption=false]{subfig}
\usepackage{braket}
\newcommand{\ba}{\begin{eqnarray}}
\newcommand{\ea}{\end{eqnarray}}
\newcommand{\bd}{\begin{displaymath}}

\newcommand{\nn}{\nonumber \\}
\DeclareGraphicsExtensions{.pdf,.png,.jpg}

\begin{document}
\title{A Subsystem Ginzburg-Landau and SPT Orders Co-existing on a Graph}

\author{Jintae Kim}
\affiliation{Department of Physics, Sungkyunkwan University, Suwon 16419, Korea}
\author{Hyun-Yong Lee}
\affiliation{Department of Applied Physics, Graduate School, Korea University, Sejong 30019, Korea}
\affiliation{Division of Display and Semiconductor Physics, Korea University, Sejong 30019, Korea}
\author{Jung Hoon Han}
\email[Electronic address:$~~$]{hanjemme@gmail.com}
\affiliation{Department of Physics, Sungkyunkwan University, Suwon 16419, Korea}
\date{\today}
\begin{abstract} We analyze a model demonstrating the co-existence of subsystem symmetry breaking (SSB) and symmetry-protected topological (SPT) order, or subsystem LSPT order for short. Its mathematical origin is the existence of both a subsystem and a local operator, both of which commute with the Hamiltonian but anti-commute between themselves. The reason for the exponential growth of the ground state degeneracy is attributed to the existence of subsystem symmetries, which allows one to define both the Landau order parameter and the SPT-like order for each independent loop. 
\end{abstract}
\maketitle

\section{Introduction}
Multiply degenerate ground states in a many-body system are usually generated as a consequence of global symmetry breaking (GSB). A prime example of GSB is the twofold degeneracy of the ground states in the Ising model, where all spins are either up or all down. One of the remarkable advances in condensed matter theory over the past decades is the identification of a new mechanism by which multiple ground state degeneracy (GSD) is generated. One of these routes is the topological order, as most dramatically realized in fractional quantum Hall systems~\cite{Laughlin87, Wen89}. Another recently discovered path to having multiple GSD is the so-called SPT order~\cite{kitaev01,wen10,wen11,wen13,cirac11}. 
In models with SPT order, the ground state is unique if defined on a closed manifold but becomes multiply degenerate on an open geometry such as the open chain. Among the tell-tale signs of SPT order is the symmetry fractionalization of the global symmetry at the edges~\cite{pollmann10,oshikawa12,turner12,nayak14} which is also responsible for the multiple GSD. A lot of different models exhibiting SPT order has been examined in the past, in particular in one dimension~\cite{quella13,chen14,motrunich14}. Some models demonstrate the GSB even for the closed geometry, together with the SPT order~\cite{quella13,motrunich14}. A recent twist to the original idea of SPT order protected by the global symmetry is the subsystem SPT (SSPT), being protected by the symmetries in a sub-manifold of the overall system~\cite{sondhi18}. 

The SPT models treated in the past considered an alternating site and link variables arranged in a one-dimensional fashion. The case of $m$ degrees of freedom residing on the vertices and $n$ degrees of freedom on the links was considered in Refs. \onlinecite{sobrinho14,sobrinho15,jimenez20,pramod20}. The model exhibited a multiple GSD even for a closed chain. When the same model was placed on a graph, the GSD increased exponentially with the number of holes, called the first Betti number, in the graph. In this paper, we give 
a clear interpretation of the results found in Ref. \onlinecite{pramod20} in terms of the notion of subsystem Landau and SPT orders. Features that are characteristic of symmetry breaking as well as SPT can be found in this model, and for each closed loop in the graph. We label such order as the subsystem Landau-SPT order, or SLSPT order for short. The observation of the co-existence of Landau and SPT orders in a given chain had been made in the past~\cite{quella13,motrunich14}, but not been extended to the case of a general graph. 
In Sec. \ref{sec:Introduction} we review the $V_2 /L_4$ model introduced in Ref. \onlinecite{pramod20}, which is a generalization of the $V_2 /L_2$ model (to be defined precisely) studied by several authors in the past~\cite{chen14,motrunich14,pramod20}. The ground state degeneracy this model is understood in terms of a properly defined Landau order parameter in Sec. \ref{sec:ground states}. The $V_2/L_4$ on a graph is analyzed in terms of the co-existing Landau and SPT orders in Sec. \ref{sec:graph}. The origin of the exponential growth of GSD is understood. We make a summary and conclude in Sec. \ref{sec:summary}. 

\section{V2/L4 model}\label{sec:Introduction} 

\begin{figure}[h]
\centering
\includegraphics[width=0.45\textwidth]{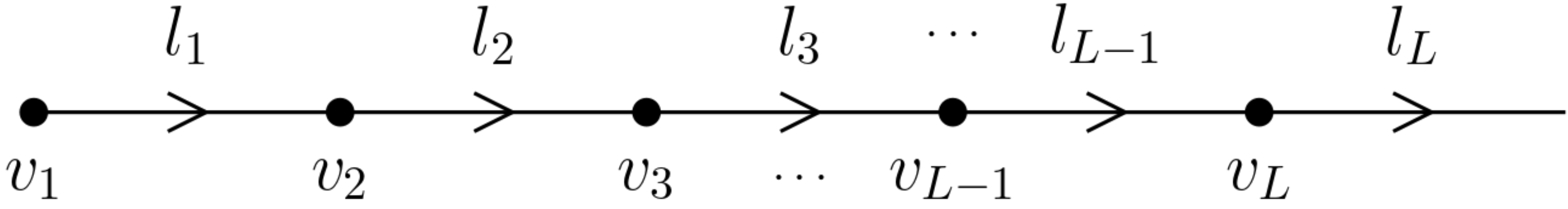} 
\caption{Schematic figure of alternating vertices and oriented links. For a closed chain the link $l_L$ attaches to the vertex $v_1$. }
\label{fig:1}
\end{figure}

For self-consistent reading of this paper, we make a brief review of the $V_2 /L_4$ model introduced in Ref. \onlinecite{sobrinho15} and analyzed in Ref. \onlinecite{pramod20}. In a family of Hamiltonians we will call the $V_m/L_n$ model, there reside $m$ degrees of freedom, or levels, at the vertices ($v$) and $n$ levels at the links ($l$). In the simplest case of a one-dimensional chain, the vertices and links appear alternatively, as illustrated in Fig. {\ref{fig:1}}. Quantum states at the vertices and links are written $|\alpha \rangle_v$ and $|\beta\rangle_l$, with $0 \le \alpha  \le m-1$ and $0 \le \beta  \le n-1$, respectively. 

The Hamiltonian of the $V_m/L_n$ model is a sum of mutually commuting projectors. One type of projector called $A_v$ is defined with respect to a vertex, while the other type $C_l$ is defined with respect to a link. Their specific expressions depend on the choice of $(m,n)$ representing the respective dimension of the Hilbert space at the vertices and links. It turns out the $m=n=2$ case had been studied extensively in the past~\cite{chen14,motrunich14}. Our focus here is on the extension to $n>2$. Striking differences from the $V_2/L_2$ situation already show up for the $V_2/L_4$ case, where we will be devoting most of our discussion.

For a one-dimensional chain the vertices and sites can be labeled as $(v_i, l_i)$, with $i$ running from 1 through $L$ for a chain of length $L$ (see Fig. \ref{fig:1}). Then we write the Hamiltonian 
\ba
H= - \sum\limits_{i=1}^L (A_i+ C_i) , \label{eq:H}
\ea
where $i = (v_i , l_i )$ stands for a combined vertex+link unit. The vertex operator $A_i$ for the $V_2 /L_4$ model acts on a given vertex $v_i$ and its two adjacent links $(l_{i-1}, l_i)$ as 
\ba
A_i &=& \frac{1}{4} \sum_{n=0}^3 (X_{l_{i-1}} x_{v_i}  X_{l_i}^3)^n \nn
& = & \frac{1}{4}(1 + X_{l_{i-1}} x_{v_i}  X_{l_i}^3 + X_{l_{i-1}}^2 X_{l_i}^2 + X_{l_{i-1}}^3 x_{v_i} X_{l_i} ) . \label{eq:A_v}\nn
\ea
The link operator $C_i$ acts on the given link $l_i$ and its two neighboring vertices $(v_i, v_{i+1})$, 
\ba C_i  =  \frac{1}{2} \sum_{n=0}^1 (z_{v_i} Z_{l_i}^2 z_{v_{i+1}} )^n
=  \frac{1}{2} ( 1 + z_{v_i} Z_{l_i}^2 z_{v_{i+1}} ) .  \label{eq:C_l} \ea
Such choice of $A$ and $C$ operators is specific to the $V_2  / L_4$ model. 

The operator form of $A_i$ depends on the direction of ``arrows" on the links. The rule is to assign $X$ to a link with an ``incoming" arrow to a vertex and $X^3$ to a link with an ``outgoing" arrow. The $A$-operator for arbitrary arrow directions on the links is therefore given by
\ba 
A_i &=& \frac{1}{4} \sum_{n=0}^3 (\prod_{l_{in}} X_{l_{in}}  \prod_{l_{out}} X_{l_{out}}^3 x_{v_i})^n . \ea
The products $\prod_{l_{in}}$ and $\prod_{l_{out}}$ mean, for example, that if both arrows are ``in", then we must write $( X_{l_{i-1}}^3 x_{v_i} X_{l_i}^3)^n$. For simplicity, we adhere to the  arrow scheme shown in Fig. \ref{fig:1} and the vertex operator definition in Eq. (\ref{eq:A_v}). 

The lower-case $x_v$ and $z_v$ operators act on the vertices. Meanwhile, upper-case $X_l$ and $Z_l$ operate on the link states. In general they satisfy $X_l|g\rangle_l = |g+1 \rangle_l$ (mod 4) and $Z_l |g\rangle = \omega^g |g\rangle_l$ with $\omega = i$, $g=0,~1,~2,~3$. Similar relations hold for the vertex operators, with $\omega = -1$. An identity
\ba Z_{l}^p X_l^q = \omega^{pq} X_l^q Z_l^p \ea 
will be used repeatedly for many of the derivations that will follow. It is easily verified that all the projectors in the Hamiltonian are mutually commuting and square to itself. 

This model we wrote down in Eq. (\ref{eq:H}) possesses some global symmetries. For the $V_2 /L_4$ Hamiltonian, there are two global symmetry operators
\ba \phi = \prod_i x_{v_i},~~\theta =  \prod_i Z_{l_i}, \label{eq:Z2Z4-symmetries} \ea
which generate the $\mathbb{Z}_2 \times \mathbb{Z}_4$ symmetry of the model. One can check that both symmetry generators commute with the Hamiltonian. A crucial aspect of the $V_2 /L_4$ model is the existence of a local $\mathbb{Z}_2$ operator $X_{l_i}^2$, which commutes with the Hamiltonian
$[H, X_{l_i}^2 ] = 0$ for any link $l_i$. Such local operator is absent in the $V_2 /L_2$ model. As a result, the eigenstates can be classified according to the set of quantum numbers or ``$p$-sectors" 
\ba p \equiv \{ p_1 , \cdots, p_L \} \ea 
where $p_i = \pm 1$ stands for the eigenvalue of $X_{l_i}^2$ at the link $l_i$. 

Within each $p$-sector one can reduce the $V_2/L_4$ model to an effective $V_2/L_2$ model. To accomplish this, first one organizes the link states in terms of the eigenstates of $X_l$:
\ba |\overline{0} \rangle_l & = &  (|0\rangle + |1\rangle_l + |2\rangle_l + |3\rangle_l ) /2 , \nn
|\overline{1} \rangle_l & = &  (|0\rangle_l + \omega^3  |1\rangle_l + \omega^2 |2\rangle_l + \omega |3\rangle_l ) /2 , \nn
|\overline{2} \rangle_l & = & (|0\rangle_l -   |1\rangle_l + |2\rangle_l - |3\rangle_l ) /2 , \nn
|\overline{3} \rangle_l & = & (|0\rangle_l + \omega |1\rangle_l + \omega^2 |2\rangle_l + \omega^3 |3\rangle_l ) /2 .  \ea
It is easily shown that
\ba X_l|\overline{n} \rangle_l & = &  \omega^n |\overline{n} \rangle_l , \nn
Z_l|\overline{n} \rangle_l & = &  |\overline{n-1}\rangle_l . \label{eq:X-and-Z} \ea
Both $|\overline{0}\rangle_{l_i} $ and $|\overline{2} \rangle_{l_i} $ share the same eigenvalue $X_{l_i}^2 = p_i = + 1$. For $|\overline{1} \rangle_{l_i} $ and $|\overline{3}\rangle_{l_i} $ the eigenvalue is $p_i = -1$. Within a given $p$-sector, we have
\ba Z_{l_i}^2 |\overline{0}\rangle_{l_i} & = & |\overline{2} \rangle_{l_i} , ~ (p_i = +1) \nn
Z^2_{l_i} |\overline{1} \rangle_{l_i} & = & |\overline{3} \rangle_{l_i} , ~ (p_i = -1) . \ea
In other words, the $Z_l^2$ operator acts effectively as a spin-1/2 Pauli operator, $Z_{l_i}^2 \rightarrow x_{l_i}$, in the two-dimensional subspace of fixed $p_i$. One can also show that $X_{l_i}$ acts as the Pauli-$z_{l_i}$ if $p_i = +1$, and as $X_{l_i} \equiv \omega z_{l_i}$ if $p_{i} = -1$. As a result, the following replacements are allowed in a given $p$-sector:
\ba X_l & \rightarrow & \omega^{ (1-p_l )/2} z_l, \nn
X_l^2 &\rightarrow & p_l, \nn
X_l^3 & \rightarrow & p_l \omega^{(1-p_l )/2} z_l . \ea
With these considerations one can reduce the $A_i$ and $C_i$ operators in Eqs. (\ref{eq:A_v}) and (\ref{eq:C_l}) as
\ba A_i & \rightarrow & A_i^{(p) } =  \frac{1}{4}\Bigl[ 1 + p_{i-1} p_{i}  \nn
&& ~~~~ +  ( p_{i-1} + p_{i} )  \omega^{- ( p_{i-1} +  p_{i} ) /2 +1}  z_{l_{i-1}} x_{v_i}  z_{l_i} \Bigr]  \nn
C_i & \rightarrow & C_i^{(p)} = \frac{1}{2} ( 1 + z_{v_i} x_{l_i} z_{v_{i+1}} )  . \label{eq:reduced-A-and-C} \ea 
Each $p$-sector then gives rise to an effective $V_2 /L_2$ Hamiltonian
\ba H^{(p)} = -\sum_i ( A_i^{(p)} + C_i^{(p)} )  \ea
with the $A_i^{(p)}$ and $C_i^{(p)}$ terms given in Eq. (\ref{eq:reduced-A-and-C}). One can view the original $V_2 /L_4$ model as the direct sum
\ba H = \oplus_p H^{(p)} , \ea
where each $H^{(p)}$ is defined in the $V_2/L_2$ subspace. From Eq. (\ref{eq:reduced-A-and-C}) we obtain identical operators $A^{(p)}_i$ when all the $p_i$'s are reversed, $p_i \rightarrow -p_i$, implying that each $V_2 /L_2$ sector ought to be {\it doubly degenerate}. The two degenerate subspaces are connected by the global operation $\prod_i Z_{l_i}$ or $\prod_i Z_{l_i}^3$, both of which implement $p_i \rightarrow -p_i$. In other words, 
\ba H^{(-p) } & = &  ( \prod_i Z_{l_i} )  H^{(p)} ( \prod_i Z_{l_i}^3 ) \nn
& = &  ( \prod_i Z^3_{l_i} )  H^{(p)} ( \prod_i Z_{l_i} ) . \ea

In particular, the ground states of the $V_2 /L_4$ model comes from the sector $p = \{ 1, \cdots , 1\}$ and $p = \{ -1, \cdots , -1\}$ where one can write $A_i$ and $C_i$ as
\ba A_i & = & \frac{1}{2}(1  +  z_{l_{i-1}} x_{v_i}  z_{l_i}  ) , \nn
C_i & = & \frac{1}{2} ( 1 + z_{v_i} x_{l_i} z_{v_{i+1}} )  . \label{eq:AC-in-V2L2} \ea
In fact, this $V_2 /L_2$ model has been studied extensively as a model for one-dimensional SPT~\cite{chen14,motrunich14}. The $V_2 /L_4$ model is a natural extension of the $V_2/L_2$ model. The pure $V_2/L_2$ model, with all $p_l = 1$ or $-1$, has the global $\mathbb{Z}_2 \times \mathbb{Z}_2$ symmetry generated by
\ba \prod_i x_{v_i} ~~ {\rm and} ~~ \prod_i x_{l_i} , \ea
but no extra local symmetry. 

\section{ground states of V2/L4 model}
\label{sec:ground states}

There are two ways to go about writing down the ground states of the $V_2/L_4$ Hamiltonian, Eq. (\ref{eq:H}). The first one is to identify the ground states of $-\sum_i C_i$ and then act on them with the projector $\prod_i A_{i}$.
The other way is to first identify the ground states of $-\sum_i A_{i}$, and then act on them with the projector $\prod_i C_i$. It is not hard to see that both ways lead to states with the eigenvalues of $A_i$ and $C_i$ all equal to +1, which by definition gives the ground states of the projector Hamiltonian. 

Given our analysis in the previous section, it seems more enlightening the analyze the ground states in the eigenbasis of $X_{l_i}^2$, where the operator $A_i$ is diagonalized easily. We obtain the two degenerate ground states of the $V_2/L_4$ model as
\ba |G_1\rangle & = & P_C |S_1 \rangle , ~~ |G_2 \rangle = P_C |S_2 \rangle \nn
|S_1 \rangle & = & ( \otimes_i |\overline{0} \rangle_{v_i})  (\otimes_j |\overline{0} \rangle_{l_j})   \nn
|S_2 \rangle & = &( \otimes_i |\overline{0} \rangle_{v_i})  (\otimes_j |\overline{1} \rangle_{l_j}) .  \label{eq:G1G2-closed-chain} \ea
The vertex eigenstates $|\overline{0}\rangle_{v_i} = ( |0 \rangle_{v_i} + |1\rangle_{v_i} ) /\sqrt{2}$ and $|\overline{1}\rangle_{v_i} = ( |0 \rangle_{v_i} - |1\rangle_{v_i} ) /\sqrt{2}$ diagonalize the $x_v$ operator. The ground states of the $V_2/L_4$ model are obtained as the projection by $P_C = \prod_i C_{i}$ on the two ``seed states" $|S_1\rangle$ and $|S_2 \rangle$. The (unique) ground state of the $V_2/L_2$ model is obtained from $|G_1\rangle$ above, by rewriting $C_i$ as in Eq. (\ref{eq:AC-in-V2L2}). The two ground states of the $V_2/L_4$ model share the properties 
\ba ( \prod_i Z_{l_i} ) |G_1 \rangle &=& |G_2 \rangle , \nn
X_{l_i}^2 |G_1 \rangle &=& + |G_1 \rangle \nn
X_{l_i}^2 |G_2 \rangle &=& - |G_2 \rangle . \label{eq:G1G2} \ea

We also show how to write down the ground states in the $Z$-basis, where the $C_i$'s are diagonalized first and $P_A = \prod_i A_{i}$ acts as a projector. The two ground states are 
\ba |G'_1 \rangle & = & P_A \left[ ( \otimes_{i}\ket{0}_{v_i} ) ( \otimes_j\ket{0}_{l_j} ) \right] , \nn
|G'_2 \rangle & = & P_A \left[ ( \otimes_i\ket{0}_{v_i} ) ( \otimes_{j\neq j'}\ket{0}_{l_j}  ) \otimes \ket{2}_{l_{j'}} \right]\ea
where $j'$ is arbitrary. These ground states share the properties
\ba X_{l_i}^2 |G'_1 \rangle &=& |G'_2 \rangle , \nn
(\prod_i Z_{l_i} ) |G'_1 \rangle &=& + |G'_1 \rangle , \nn
(\prod_i Z_{l_i} ) |G'_2 \rangle & = & - |G'_2 \rangle . \label{eq:GS2} \ea
Comparing Eqs. (\ref{eq:G1G2}) and (\ref{eq:GS2}), one concludes 
\ba |G_1 \rangle & = &  (|G'_1 \rangle + |G'_2 \rangle )/\sqrt{2} \nn
|G_2 \rangle & = & ( |G'_1 \rangle - |G'_2 \rangle  )/\sqrt{2}  . \ea


In what follows, we provide a geometrical interpretation of the ground state $|G_1\rangle$. Due to $z_v^2=1$, the product of neighboring $z_{v_i} Z_{l_i}^2 z_{v_{i+1}}$ is simply $Q_{12} = z_{v_1} \left(  \prod_{i_\mathcal{S}} Z^2_{l_i}\right) z_{v_2}$ where $i_\mathcal{S}$ stands for the links between the left-most vertex $v_1$ and the right-most one $v_2$. Its action on the seed state $|S_1\rangle$ permutes the link states along $\mathcal{S}$, i.e., $|\overline{0}\rangle_l \rightarrow |\overline{2}\rangle_l$ and flips the vertex states at edges $|\overline{0}\rangle_{v_{1/2}} \rightarrow |\overline{1}\rangle_{v_{1/2}}$, or graphically illustrated as

\begin{align}
 \includegraphics[width=0.15\textwidth]{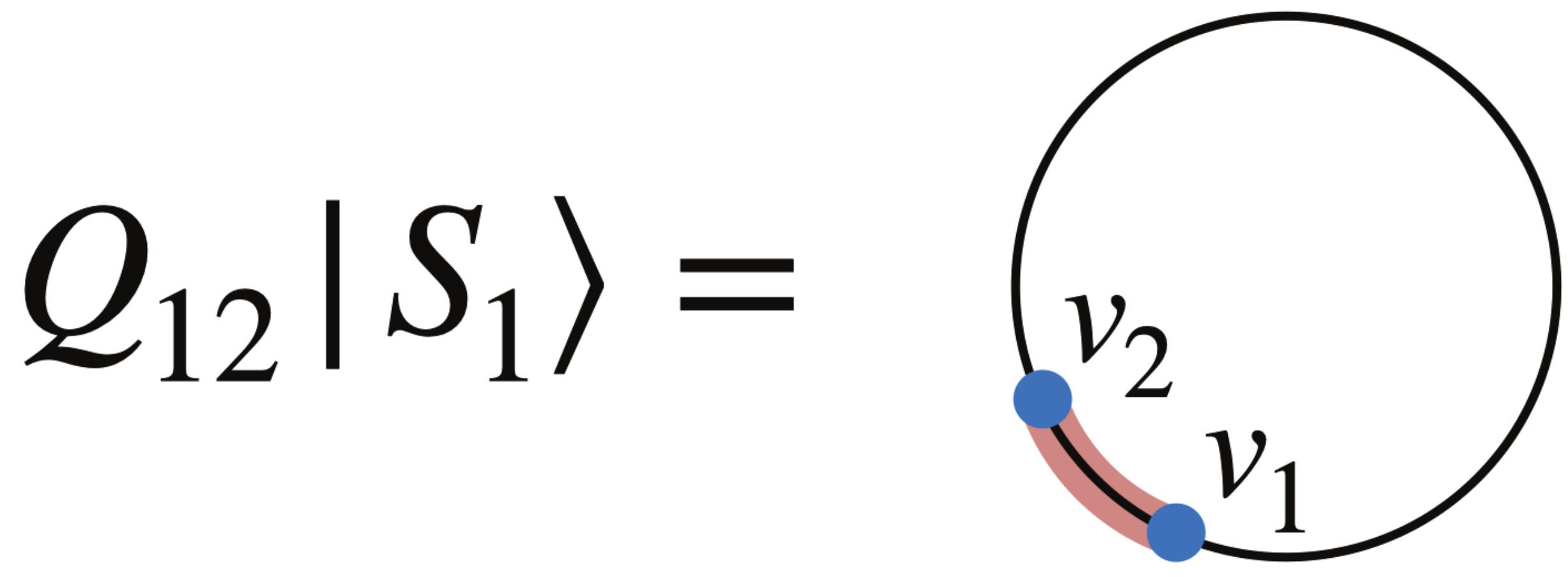},\nonumber
\end{align}
where blue dot stands for each edge vertex while the red string for all links and vertices along $\mathcal{S}$. Consequently, the expansion of the projector $P_C$ leads to the super position of all possible string configurations on the circle:

\begin{align}
 \includegraphics[width=0.43\textwidth]{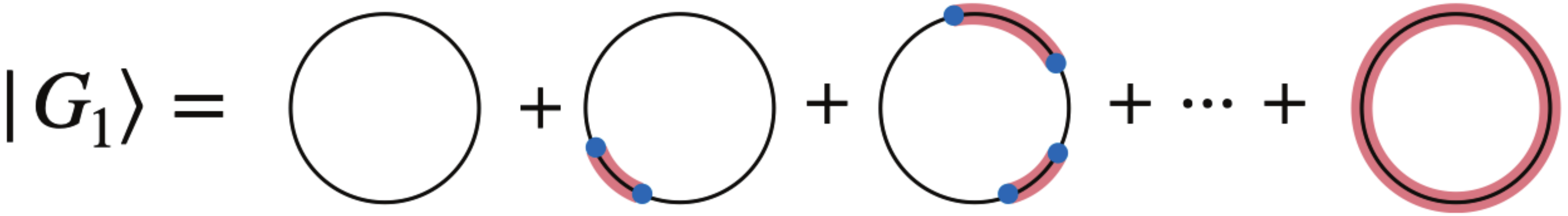}. 
\end{align}
Due to $|G_2\rangle = \prod_i Z_{l_i} |G_2\rangle$, the geometrical interpretation of $|G_2\rangle$ is given in the same manner. Only difference is that all link states are raised by one, i.e., $|\overline{n}\rangle_l \rightarrow |\overline{n+1}\rangle_l$. 

In an open chain of the same model, some edge states appear as a consequence of SPT, as thoroughly analyzed in Ref. \onlinecite{pramod20}.

\section{V2/L4 Model on a graph}
\label{sec:graph}

\subsection{GSD on a Graph}

\begin{figure}[h]
\centering
\includegraphics[width=0.47\textwidth]{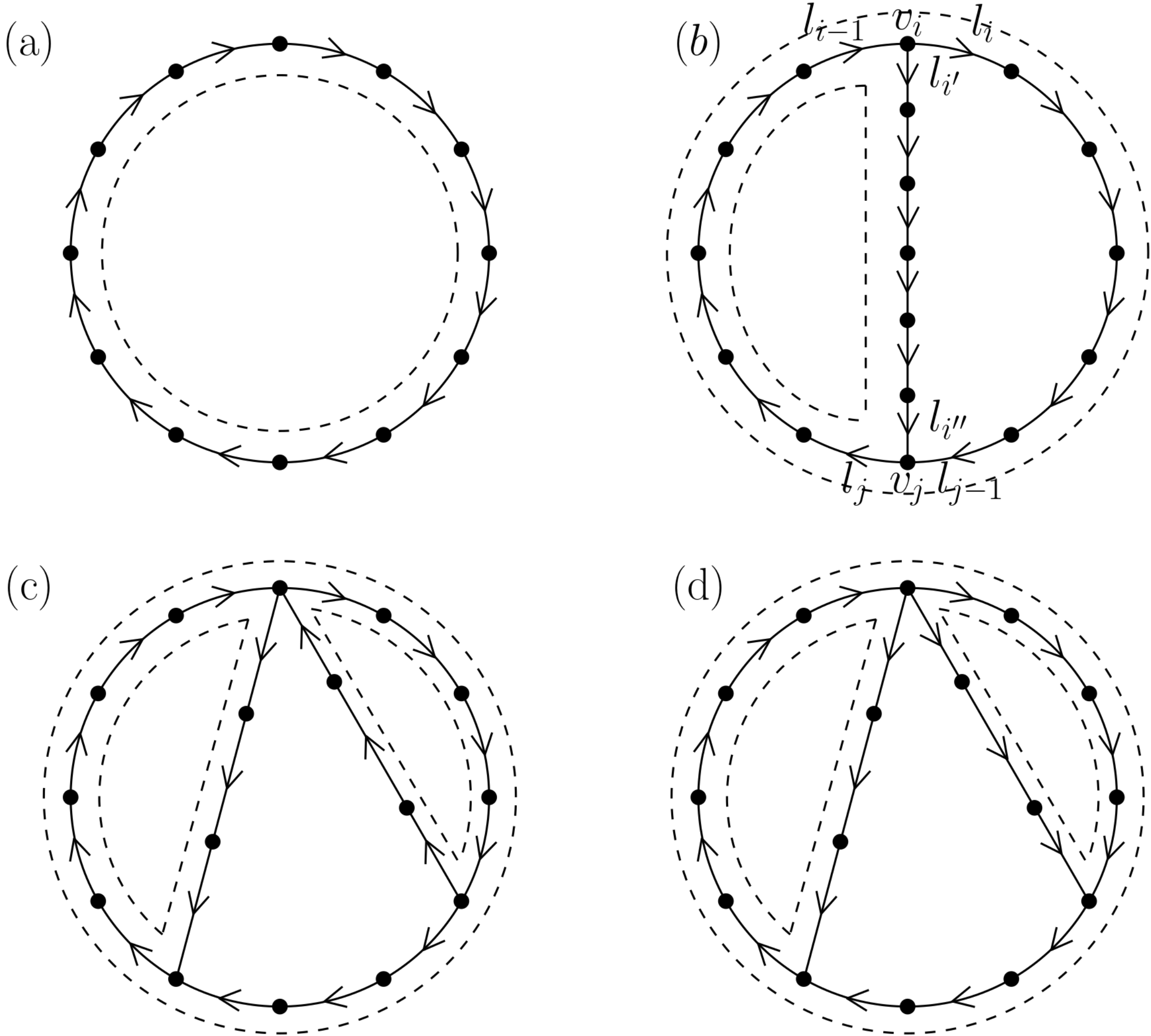} 
\caption{Examples of closed and connected planar graphs with various Betti numbers (a) $B_1 = 1$, (b) $B_1 = 2$, and (c), (d) $B_1 = 3$. Dots are the vertices and the arrows represent the links of the graph. Dashed lines are the independent closed loops where we can perform a subsystem $\mathbb{Z}_4$ symmetry transformation. Because of the orientation, the loop operators correspond to the right semi-circular loop are different for (c) and (d).}
\label{fig:2}
\end{figure}

The discussion of the ground states of the $V_2/L_4$ model both in the closed and the open chain thus far might suggest that we are merely dealing with what seems to be two copies of the  well-known $V_2/L_2$ model. Interestingly, the real point of departure between the two families of models occurs when these models are put on a {\it graph}\cite{pramod20}. The closed circle is a simplest example of a graph with the first Betti number $B_1 = 1$. Now, one can imagine putting the model on a more intricate graph such as shown in Fig. \ref{fig:2}(b), which has $B_1 = 2$. Intuitively, the first Betti number measures the number of {\it cycles} or independent loops in a given graph. 

Considering a $B_1 = 2$ graph such as shown in Fig. \ref{fig:2}(b), one can write the same $A_i$ and $C_i$ projectors as in the $B_1 = 1$ graph, namely a circle, except at the two vertices $v_i$ and $v_j$ where two lines become connected at the vertex. At these vertices, there are three (not two) links which are connected to a single vertex. The definition of the vertex operators which are projectors and commute with other vertex operators must generalize accordingly, 
\ba 
A_i &=& \frac{1}{4}(1 \! + \! X_{l_{i-1}}  X_{l_i}^3 X_{l_{i'}}^3 x_{v_i} \!\nn
&&+\! X_{l_{i-1}}^2 X_{l_i}^2 X_{l_{i'}}^2\!+\! X_{l_{i-1}}^3  X_{l_i} X_{l_{i'}} x_{v_i})\nn
A_j &=& \frac{1}{4}(1 \! + \! X_{l_{j-1}} X_{l_j}^3 X_{l_{i''}} x_{v_j}  \!\nn
&&+\! X_{l_{j-1}}^2 X_{l_j}^2 X_{l_{i''}}^2\!+\! X_{l_{j-1}}^3  X_{l_j} X_{l_{i''}}^3 x_{v_j}) \label{eq:A-at-junction} 
\ea
where $v_i$ ($v_j$) has one (two) arrowhead towards it and two (one) arrowheads away from it. In fact, a completely general definition of the vertex operator for arbitrary graph is possible as
\ba A_i = \frac{1}{4} \sum_{j=0}^{3} \left[ (\prod_{q \in in}  X_{l_{q,i}} \prod_{q' \in out} X_{l_{q' ,i}}^{3}) x_{v_i}  \right]^j  .  \label{eq:Av-general} \ea
There is a factor $X$ ($X^3$) for the links whose arrows come into (out of) the vertex. The link operator $C_i$ is the same as in Eq. (\ref{eq:C_l}) regardless of the graph type. Despite the much complex forms, $A_i$ and $C_j$ always remain mutually commuting projectors.  The $V_2 /L_4$ Hamiltonian on an arbitrary graph generalizes accordingly, 
\ba H = -\sum_{i=1}^{N_v} A_i-\sum_{j=1}^{N_l} C_j , \ea
spanning all the vertices and the links in the graph and using Eq. (\ref{eq:Av-general}) for the vertex operator. The number of vertices ($N_v$) and of links ($N_l$) are no longer equal for a general graph but are rather related by the first Betti number,
\ba N_l - N_v + 1 = B_1 . \ea


\subsection{Subsystem Symmetries}

We can discuss how to explicitly construct the multitude of degenerate ground states on a graph. For example, the $V_2/L_4$ model defined on a graph shown in Fig. \ref{fig:2}(b) allows two loop operators that commute with the Hamiltonian. We call them $\theta_1$ and $\theta_2$, and they consist of the product of $Z_l$'s along the left semi-circular loop and the large circular loop, respectively. These two loops are drawn as dashed lines in Fig. \ref{fig:2}(b). Note that the definition of these loop operators are obtained by simply ``following the arrows" drawn on the graph. On a $B_1 = 2$ graph we have two such independent loops. Accordingly one can write down four independent ground states, which are $|G_1\rangle = P_C \left[( \otimes_i |\overline{0} \rangle_{v_i})  (\otimes_j |\overline{0} \rangle_{l_j})\right] $, and
\ba |G_2 \rangle & = & \theta_1 |G_1 \rangle, \nn
|G_3 \rangle & =  & \theta_2 |G_1 \rangle , \nn
|G_4 \rangle & = & \theta_1 \theta_2 |G_1 \rangle . \ea
At first sight the construction of degenerate ground states on a graph bears resemblance to the way that topologically distinct states are generated on a finite-genus space. The GSD formula ${\rm GSD} = 2^{B_1}$ has resemblance to the formula for the topological ${\rm GSD} = 2^{2g}$, which applies to {\it two-dimensional} topological models such as the toric code, where $g$ (also known as the second Betti number) is the genus of two-dimensional surface. 

\begin{figure}[ht]
\centering
\includegraphics[width=0.47\textwidth]{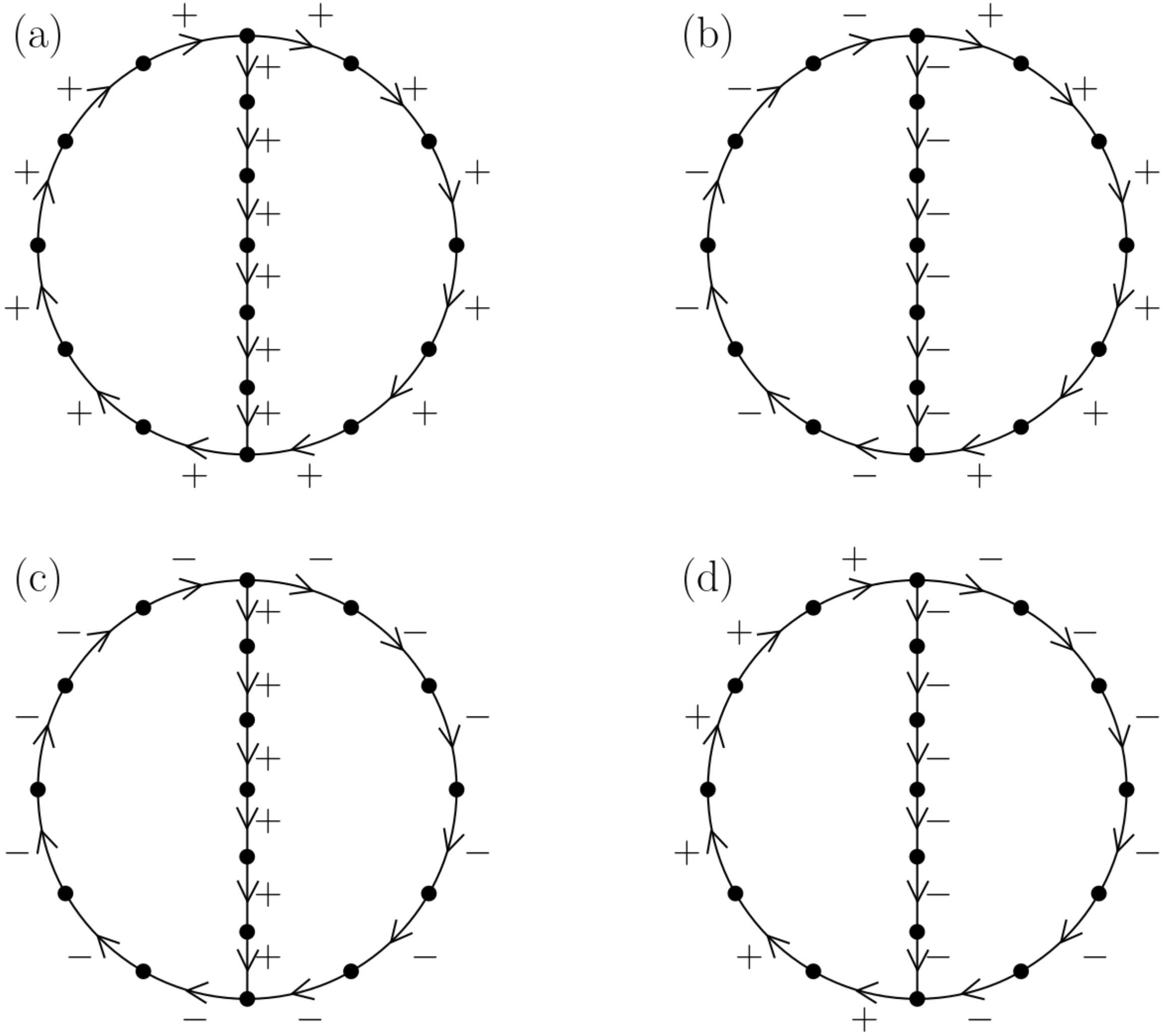} 
\caption{Four kinds of configurations of $p$'s that reduce the $V_2/L_4$ model to the $V_2/L_2$ model when $B_1=2$. The $\pm$ signs on the links represent the eigenvalues of $X_l^2$, or equivalently, the $p_l$'s. }
\label{fig:3}
\end{figure}

Unlike the topologically ordered states which do not have local order parameters, the four ground states derived above can be distinguished by their ``order parameter" $X_l^2$. With $|G_1\rangle$, the expectation value is $\langle G_1 | X_l^2 |G_1 \rangle = +1$ on every link of the graph. For $|G_2\rangle$, the links along the loop where $\theta_1$ acts have $\langle X_l^2 \rangle = -1$. For $|G_3\rangle$, it is the links along the outer perimeter where the order parameters are reversed. Finally in the fourth ground state $|G_4 \rangle$ it is the other inner loop where the links have $\langle X_l^2 \rangle = -1$. The four order parameter patterns are depicted in Fig. \ref{fig:3}. There is a clear parallel to the usual classification of states by order parameters, but one must carefully note that its nature is not entirely global. Perhaps a better termnology is the {\it subsystem symmetry breaking} (SSB, not to be confused with the {\it spontaneous symmetry breaking}) and distinguish it from the global symmetry breaking (GSB) of most many-body models.

Along a similar line of reasoning, the loop operator consisting of the product of $Z_l$'s and $Z_l^3$'s along the independent loops as dictated by the flow of arrows commute with the Hamiltonian and performs the {\it subsystem $\mathbb{Z}_4$ symmetry transformation}. In the case of Fig. \ref{fig:2}(c), all such loop operators consist of the product of $Z_l$'s only. On the other hand, the loop operator corresponding to the right semi-circuplar loop in Fig. \ref{fig:2}(d) is $\prod_i Z_{l_i} \prod_{i'} Z_{l_{i'}}^3$ where $i$'s are the links on the arc and $i'$'s are the links on the inner segment. The number of independent loops equals the Betti number $B_1$, and the degenerate ground states are generated by applying the subsystem loop operators and their products on one particular ground state $| G\rangle$. There are exactly $2^{B_1} -1$ different products of subsystem loop operators available, for a total of $2^{B_1}$ ground states on a graph with the Betti number $B_1$.

The $V_2/L_4$ Hamiltonian on a graph still has the exact symmetry $[H , X_l^2] = 0$ for all the links $l$. As a result, $X_l^2$ operators in the graph model can be replaced by their respective quantum numbers $p_l$. In particular, the vertex operators at the junction given in Eq. (\ref{eq:A-at-junction}) become, in a given $p$-sector,
\ba 
A_i^{(p)} &=& \frac{1}{4} \bigl[ 1 \! + p_{i-1} p_{i} p_{i'}  \nn
&+ & ( p_{i} p_{i'}  \!+\! p_{i-1} ) \omega^{(3-p_{i-1} - p_{i} - p_{i'}) /2 }   z_{l_{i-1}}  z_{l_i} z_{l_{i'}} x_{v_i} \bigr] \nn
A_j^{(p)} &=& \frac{1}{4} \bigl[ 1 \! + p_{j-1} p_{j} p_{i'}  \nn
&+ & ( p_{j-1} p_{i'}  \!+\! p_{j} ) \omega^{(3-p_{j-1} - p_{j} - p_{j'}) /2 }   z_{l_{j-1}}  z_{l_j} z_{l_{i'}} x_{v_i} \bigr]. \nn
\ea
One can show, by explicit calculation, that $A_i^{(p)}$ becomes $\frac{1}{2} \bigl( 1 +z_{l_{i-1}}  z_{l_i} z_{l_{i'}} x_{v_i} \bigr)$ for $\{p_{i-1},p_i,p_{i'}\}$ $=\{1,1,1\}$, $\{-1,1,-1\}$, $\{-1,-1,1\}$. In a similar manner, we have $A_j^{(p)}=\frac{1}{2} \bigl( 1 +z_{l_{j-1}}  z_{l_j} z_{l_{i'}} x_{v_j} \bigr)$ for $\{p_{j-1},p_j,p_{i'}\}=$ $\{1,1,1\}$, $\{-1,-1,1\}$, $\{1,-1,-1\}$. The $V_2/L_2$ model on a graph comes from having the following choice of vertex operators: 
\ba
A_i^{(p)}&=&\frac{1}{2} \bigl( 1 +z_{l_{i-1}}  z_{l_i} z_{l_{i'}} x_{v_i} \bigr)\nn
A_j^{(p)}&=&\frac{1}{2} \bigl( 1 +z_{l_{j-1}}  z_{l_j} z_{l_{i'}} x_{v_j} \bigr) \label{eq:V2L2g}
\ea
at the two junctions $i$ and $j$ shown in Fig. \ref{fig:2}(b). For all other vertices and all the links one has the usual definition of the vertex and link operators given in Eq. (\ref{eq:AC-in-V2L2}). After some enumeration, one finds the four configurations of $p$'s shown in Fig. \ref{fig:3} can reduce the $V_2/L_4$ model to the $V_2/L_2$ model on the graph with $B_1 = 2$. This argument again shows why there is fourfold degeneracy of the ground states on the $B_1 = 2$ graph.

It is worth examining the general character of the pure $V_2 /L_2$ model on a graph. It can be shown that GSD of the $V_2/L_2$ model remains at GSD=1 regardless of the Betti number of the graph. The global $\mathbb{Z}_2 \times \mathbb{Z}_2$ symmetry of the $V_2/L_2$ model on a simply connected graph with $B_1 = 1$ is partially lost due to the vertex terms at the junction, Eq. (\ref{eq:V2L2g}). One can prove quite easily that although $\prod_i x_{v_i}$ remains a symmetry, $\prod_i x_{l_i}$ no longer commutes with the vertex operators at the junction and hence fails to be a symmetry operator. The global symmetry of the $V_2 /L_2$ model is  lowered from $\mathbb{Z}_2 \times \mathbb{Z}_2$ to $\mathbb{Z}_2$ on a multiply connected graph. On further observation, however, we realize that a {\it partial} product of link operators $\prod'_i x_{l_i}$ for the links forming a closed loop {\it does commute} with the junction terms in Eq. (\ref{eq:V2L2g}) and restore the $\mathbb{Z}_2 \times \mathbb{Z}_2$ symmetry for that loop. Contrasted with the usual SPT, this is a realization of the subsystem SPT, or SSPT~\cite{sondhi18}. By implication, when we cut open any segment of the graph, the emerging edge behavior and symmetry fractionalization will be exactly those of the open chain case already analyzed. Since the ground states of the pure $V_2/L_2$ model are ground states of the $V_2/L_4$ model, even in $V_2/L_4$ model there is SSPT.

\section{Excitations}
\label{sec:excitation} 

The concept of excitation arises naturally in frustration-free models to which our $V_2/L_4$ model belongs. The ground state(s) has all of the eigenvalues of $A_i$ and $C_i$ equal to +1, and excited states should have one of these equal to zero instead~\cite{pramod20}. Depending on whether the link or the vertex operator eigenvalues change from 1 to 0, one can make a distinction between link excitations and vertex excitations. Another useful way to classify excitations is in terms of the changes in the $p$-eigenvalues, $p = \{ p_1, \cdots, p_L \}$, of the chain. Changes in any of the eigenvalues in the $p$-set leads to different sectors of the block, $H = \oplus_p H^{(p)}$, where $H$ is the original $V_2/L_4$ Hamiltonian and each $H^{(p)}$ represents some realization of the $V_2/L_2$ Hamiltonian. We will examine the nature of excitations from both perspectives. Eigenvalues of $A_i$ and $C_i$ will be denoted $a_i$ and $c_i$ from now on. 

The link excitation is attained by rewriting one of the operators $C_i$ in the projector $P_C$ by its orthogonal complement~\cite{pramod20}
\ba C_i^\perp = \frac{1}{2} (1- z_{v_i} Z_{l_i}^2 z_{v_{i+1}} ) , ~~ C_i^\perp C_i =0 .  \ea
We can define the projector
\ba P_C (j) = (\prod_{i < j} C_i ) C_j^\perp (\prod_{i>j} C_i) ,\ea
and use such projector to create a link-excited state
\ba |l_j \rangle = P_C (j) |S\rangle \ea
with $|S\rangle$ being one of the two seed states. Since $C_j |l_j \rangle = 0$, one concludes $c_j = 0$, qualifying it as a link excitation. Note that $X_{l_j} C_j = C_j^\perp X_{l_j}$, and therefore
\ba X_{l_j} |G \rangle &=& X_{l_j}   P_C |S \rangle = P_C (j) X_{l_j} |S\rangle \nn
& = &  \pm P_C (j) |S\rangle  = \pm |l_j \rangle , \label{eq:X-excitation} \ea
where the $\pm$ sign comes from having $|S\rangle = |S_1\rangle$ or $|S\rangle = |S_2 \rangle$. In other words, link excitations are created by applying $X_{l_i}$ on a given link to the ground states. By a similar consideration, we learn that $x_{v_j}$ acting on the ground states generate a link-pair excitation:
\ba x_{v_j} |G \rangle = |l_{j-1}, l_{j} \rangle . \ea
The proof comes from the observation, 
\ba x_{v_j} P_C = ( \prod_{i < j-1} C_i ) C_{j-1}^\perp C_j^\perp ( \prod_{i> j+1} C_i ) . \ea
A string of $x$-excitations acting on the ground states gives
\ba
x_{v_2}\cdots x_{v_k}\ket{G}&=&\ket{l_{1},l_k}
\ea
%
and when the string forms a closed loop, the state gets back to the original ground state, e.g. $\prod_{i\in loop} x_{v_i}\ket{G}=\ket{G}$.

Both types of link-excited states $X_{l_j} |G\rangle$ and $x_{v_j} |G\rangle$ share the same $p$-eigenvalues as the ground state. This is seen by the fact that the operator that defines the $p$-eigenvalue, $X_{l}^2$, commutes with with $X_l$ and $x_v$ which create the link excitations. We conclude that the link excitations $X_{l_j} |G\rangle$ and $x_{v_j} |G\rangle$ occurs in the same $p$-sector as the ground state (intra-$p$ excitation). In this regard, the link excitations we write down here are the same as what the $V_2/L_2$ model would give.

The $p$-altering or inter-$p$ excitations are created by acting with $Z_{l_j}$ directly on the seed states. Since $Z_{l_i}$ commutes with the projector $P_C$, the $p$-altering excited state is given by
\ba P_C ( Z_{l_j} |S\rangle ) = Z_{l_j} |G \rangle . \ea
One can easily check that $X_{l_j}^2$ anti-commutes with $Z_{l_j}$, therefore $p_{l_j}$ changes from +1 to -1. This state, interestingly, continues to remain an eigenstate of $C_i$ with $c_i = +1$ everywhere. Contrary to naive expectation, $Z_{l_j} |G\rangle$ is {\it not} a link excitation. When one applies $A_i$ to this state, some nontrivial changes are discovered:
\ba A_{j} Z_{l_j} &=& \frac{1}{4} Z_{l_j} (1 + \omega^3 X_{l_{j-1}} x_{v_j} X_{l_j}^3 \nn
&& + \omega^2 X_{l_{j-1}}^2 X_{l_j}^2 + \omega X_{l_{j-1}}^3 x_{v_j} X_{l_j}) \nn
A_{j+1} Z_{l_j} &=& \frac{1}{4} Z_{l_j} (1 + \omega X_{l_j } x_{v_{j+1}} X_{l_{j+1}}^3 \nn
&& + \omega^2 X_{l_j}^2 X_{l_{j+1}}^2 + \omega^3 X_{l_j}^3 x_{v_{j+1}} X_{l_{j+1}}) . \label{eq:AZ-commute} \ea 
%
The appearance of new terms on the right side of the equations necessitates that we define some new vertex projectors as follows: 
\ba A_i (n ) &=&  \frac{1}{4}(1 + \omega^n X_{l_{i-1}} x_{v_i}  X_{l_i}^3\nn
&&~~ + \omega^{2n} X_{l_{i-1}}^2 X_{l_i}^2 + \omega^{3n} X_{l_{i-1}}^3 x_{v_i} X_{l_i} ) . \label{eq:A_v-full} \ea
The vertex projector used to define the Hamiltonian is recovered for $n=0$, and the relations derived in Eq. (\ref{eq:AZ-commute}) are now succinctly summed up:
\ba A_{j} Z_{l_j} &=& Z_{l_j} A_j (3) \nn
A_{j+1} Z_{l_j} &=& Z_{l_j} A_{j+1}  (1) . \ea 
%
The $A_i (n)$ projectors commute with $C_j$ and $A_i (n) A_i (n') = 0$ unless $n = n'$. Furthermore, $A_i (n)|S\rangle = 0$ unless $n=0$. Due to these properties, we can show 
%
\ba A_j  Z_{l_j} |G\rangle =  A_{j+1} Z_{l_j} |G \rangle = 0 . \ea
%
In other words, a pair of vertex excitations with $a_j = a_{j+1} = 0$ has been created. We conclude that different $p$-sectors of the $V_2/L_4$ model are connected through the creation of vertex-pair excitations. A $p_{l_j} = 1 \rightarrow -1$ implies that a pair of adjacent vertices have been excited, $a_j = a_{j+1} = 0$. We may sum up the situation as
%
\ba 
Z_{l_j} |G\rangle  & = &  |\omega^3_{v_j} , \omega_{v_{j+1}} \rangle , \nn
Z^3_{l_j} |G\rangle & = &  |\omega_{v_j} , \omega^3_{v_{j+1}} \rangle . \label{eq:vertex-pair-excitation} \ea
%
The symbol on the right $|\omega^n_v \rangle$ means that the seed state has been acted on by $A_v (n)$ before the projector $P_C$ is applied. The vertex excitations carrying $\omega^3$ and $\omega$ as quantum numbers must exist as a pair, and serve to connect different $p$-sectors. 

On the other hand, exciting the vertex state by $z_{v}$ or a link state by $Z_l^2$ does not change the $p$-eigenvalues, and leads to the following excitations: 
%
\ba z_{v_j} |G\rangle  & = & | \omega^2_{v_j} \rangle , \nn
%
Z_{l_j}^2 |G\rangle & = & |\omega^2_{v_j}, \omega^2_{v_{j+1}} \rangle =z_{v_j} z_{v_{j+1}}  |G\rangle . \label{eq:z-excitation} 
\ea
The vertex excitation with  the quantum number $\omega^2 = -1$ can exist in isolation. The link projector $C_i$ has the useful identity $C_j = z_{v_j} z_{v_{j+1}} Z_{l_j}^2 C_j$, which one can verify directly from its definition. As a result, $P_C$ acting on some seed state $|S\rangle$ leads to the same consequence as when acting on a different seed state $z_{v_j} z_{v_{j+1}} Z_{l_j}^2 |S\rangle$. This leads to the identity mentioned in the second equation above. 

The $p$-changing vertex-pair excitation created by $Z_{l_j}|G\rangle$ can be generalized by considering a string of $Z$-excitations given by the product
%
\ba Z_{l_1} \cdots Z_{l_k} |G \rangle = |\omega_{v_1}^3 , \omega_{v_{k+1}} \rangle  .\ea
By the time the product $\prod_i Z_{l_i}$ forms a closed loop, one reaches the other ground state, e.g. $( \prod_{i \in loop} Z_{l_i} )  |G_1 \rangle = |G_2 \rangle$. To sum up, there are intra-$p$ excitations in the form of link excitations [Eq.  (\ref{eq:X-excitation})] and vertex excitations [Eq. (\ref{eq:z-excitation})], and inter-$p$ excitations in the form of vertex-pair excitations [Eq. (\ref{eq:vertex-pair-excitation})]. This gives the complete classification of the elementary excitations in the $V_2 /L_4$ model. 

\section{Discussion} \label{sec:summary} 
In this paper we have analyzed the properties of the $V_2 /L_4$ model on a general graph~\cite{pramod20}.  
The Hilbert space of the model is block-diagonalized by a set of local quantum numbers $\{ p_{l_i}  = \pm 1 \}$, and we have shown that the $V_2/L_4$ model maps to a general $V_2/L_2$ model in each $p$-sector. 

The GSD of our model grows exponentially with the Betti number $B_1$ characterizing the number of cycles in a graph. In fact one can easily show that the exponential growth of GSD with the Betti number is not unique to the $V_2 /L_4$ model. Even a simple Potts model on a graph can be defined in a way that exhibits the same GSD behavior. For that, one considers the same kind of graph as before and place $n$ degrees of freedom at each link labeled by the variable $z_l =0, \cdots, n-1$, and none on the vertices. We then have the Potts interaction $- \delta_{z_l,z_{l'}}$ for the neighboring links $(l, l')$, except when more than two lines meet at a vertex. In that case we have the three-link interaction $-\delta_{z_{l_1},z_{l_2},z_{l_3}}$ among the three links $(l_1 , l_2 , l_3)$ joined at a vertex. For more than three links, one simply takes the delta function of all the links $-\delta_{z_{l_1}, z_{l_2}, \cdots}$ with an overall minus sign. It is easily verified that the ground states have the $z_l$'s distributed in exactly the same pattern as those of $X_l^2$'s (or $p_l$'s) shown in Fig. \ref{fig:3} when $n=2$. For general graphs and $n$ degrees of freedom, GSD equals $n^{B_1}$, but no feature of SPT or symmetry fractionalization would be present in such models. 

To sum up, the subsystem symmetry and its breaking observed in the $V_2 /L_4$ model is not related to the SPT nature of the phase but rather co-exist with it. The idea of subsystem symmetries is applicable for both the Landau order parameter characterized by $X_l^2$, and the SPT order. 

\acknowledgments H. J. H. was supported by the Quantum Computing Development Program (No. 2019M3E4A1080227).  H.-Y.L. was supported by a Korea University Grant and National Research Foundation of Korea\,(NRF-2020R1I1A3074769). We appreciate enlightening discussion with Munjip Park.

\bibliography{SC}
\end{document}